\documentclass[12pt]{article}
\usepackage{amsmath}
\usepackage{graphicx}

\begin{document}

\begin{center}

{\bf\large{Mass and width of an unstable particle}}\\

\bigskip

Scott Willenbrock \\

\bigskip

Department of Physics, University of Illinois at Urbana-Champaign \\
1110 West Green Street, Urbana, Illinois 61801, USA \\

\bigskip

willen@illinois.edu

\end{center}

\medskip

\noindent {\bf Abstract}: We show that the mass and width of an unstable particle are precisely defined by the pole in the complex energy plane, $\mu = m - (i/2)\Gamma$, by identifying the width, $\Gamma$, with the particle's decay rate and the mass, $m$, with the oscillatory frequency. We find that the physical $Z$ boson mass lies 26 MeV below its quoted value, while the physical $W$ boson mass lies 20 MeV below. We also clarify the various Breit-Wigner formulae that are used to describe a resonance.

\bigskip\medskip

The squared mass of a stable particle is equal to the pole in the particle's propagator. For an unstable particle the pole is complex, and is related to both the mass and the width of the particle. In this paper we make this relation precise.

For simplicity we consider a complex scalar field; the calculation also applies to the transverse part of a massive vector field, such as the $W$ and $Z$ bosons. We study the propagator, which in a free field theory is the amplitude for a particle to be created at the origin and destroyed at spacetime point $x$ (or an antiparticle to be created at $x$ and destroyed at the origin),
\begin{equation}
\langle 0 | T \phi(x) \phi^\dagger (0) |0\rangle = \int \frac{d^4p}{(2\pi)^4}\frac{ie^{-ip\cdot x}}{p^2-m^2+i\epsilon}\;.
\end{equation}
The squared mass of the particle is the pole in the momentum-space propagator.

We now include interactions to all orders in perturbation theory.
Consider the self-energy corrections to the momentum-space propagator from the diagrams in Fig.~1, where the shaded circle denotes the bare one-particle-irreducible self-energy to all orders in perturbation theory, $i\Pi(p^2)$.
Summing the series of diagrams gives
\begin{eqnarray}
&&\frac{i}{p^2 - m_B^2+i\epsilon}\left(1-\frac{\Pi(p^2)}{p^2-m_B^2+i\epsilon}+\cdots\right)\\
&&=\frac{i}{p^2 - m_B^2 + \Pi(p^2)+i\epsilon}
\label{bare}
\end{eqnarray}
where $m_B$ is the bare mass.
The pole of the full momentum-space propagator, $\mu^2$, is the solution to the equation
\begin{equation}
\mu^2-m_B^2+\Pi(\mu^2)=0\;.
\label{pole}
\end{equation}
Combining Eqs.~(\ref{bare}) and (\ref{pole}) yields
\begin{equation}
\langle 0 | T \phi_B(x) \phi_B^\dagger (0) |0\rangle = \int \frac{d^4p}{(2\pi)^4}\frac{ie^{-ip\cdot x}}{p^2-\mu^2+\Pi(p^2)-\Pi(\mu^2)+i\epsilon}
\label{bare2}
\end{equation}
As usual, we renormalize the fields such that the residue of the pole is the same as in the free field theory (see Sec.\ 10.3 of Ref.~\cite{W}).  We define the renormalized fields $\phi_R = Z^{-1/2}\phi_B$, $\phi_R^\dagger = Z^{-1/2}\phi_B^\dagger$; note that $\phi_B^\dagger$ is renormalized by the same factor as $\phi_B$, rather than by the complex-conjugate factor (see Sec.\ 4 of Ref.~\cite{DDRW}). The full propagator for the renormalized fields is thus
\begin{equation}
\langle 0 | T \phi_R(x) \phi_R^\dagger (0) |0\rangle = \int \frac{d^4p}{(2\pi)^4}\frac{iZ^{-1}e^{-ip\cdot x}}{p^2-\mu^2+\Pi(p^2)-\Pi(\mu^2)+i\epsilon}
\label{unbare2}
\end{equation}
where 
\begin{equation}
Z^{-1} = 1 + \Pi^\prime(\mu^2)\;.
\end{equation}

If the particle is unstable $\Pi(\mu^2)$ is complex, and hence, from Eq.~(\ref{pole}), $\mu^2$ is complex. The pole position is gauge invariant \cite{S,K,GG} and infrared safe \cite{K,GG}. The fact that the pole is complex is a fundamental feature of an unstable particle, and will be used in the next section to derive the mass and width.

\newpage

\noindent{\bf Width and Mass}

\bigskip

The width of an unstable state, whether it be a fundamental particle or a composite, is generally understood to be the decay rate of the particle, that is, $\Gamma = 1/\tau$, where $\tau$ is the lifetime. This provides an unambiguous definition of $\Gamma$, and is used in all branches of physics.

The scattering amplitude of a process with an intermediate unstable particle acquires a complex pole from the propagator,
\begin{equation}
S \sim \frac{1}{p^2-\mu^2}\;.
\label{LOprop}
\end{equation}
Going to the rest frame of the unstable particle (we consider a general frame in Appendix A), we can rewrite Eq.~(\ref{LOprop}) as
\begin{equation}
S \sim \frac{1}{p_0^2-\mu^2}=\frac{1}{(p_0 - \mu)(p_0 + \mu)}\;.
\label{Epole}
\end{equation}
To find the time dependence of the scattering amplitude, we Fourier transform from energy to time \cite{B}:
\begin{equation}
S \sim \int_{-\infty}^\infty dp_0\;  \frac{e^{-ip_0t}}{(p_0 - \mu)(p_0 + \mu)}\;.
\label{FT}
\end{equation}
For $t>0$ we can close the contour in the lower half complex $p_0$ plane, as shown in Fig.~2, since the integral along the large semicircle is exponentially damped. Using the residue theorem, we pick up the contribution of the pole at $p_0 = \mu$, 
\begin{equation}
S \sim e^{-i\mu t} = e^{-i{\rm Re}\,\mu t}e^{{\rm Im}\,\mu t}\;.
\label{time}
\end{equation}
The decay probability is given by the square of the scattering amplitude,
\begin{equation}
|S|^2 \sim e^{2{\rm Im}\,\mu t}\;.
\end{equation}
This corresponds to exponential decay with $2{\rm Im}\,\mu = -\Gamma$. Hence we learn that ${\rm Im}\,\mu = - \Gamma/2$.

For $t<0$ we close the contour in the upper half complex $p_0$ plane and pick up the contribution of the pole at $p_0 = -\mu$. This pole is associated with antiparticle propagation. The residue theorem gives $S \sim e^{i{\rm Re}\,\mu t}e^{-{\rm Im}\,\mu t}$, and hence $|S|^2 \sim e^{\Gamma t}$. This also corresponds to exponential decay since $t$ is negative, and proves that the particle and antiparticle have the same lifetime.

The oscillatory frequency of the state is dictated by the particle energy, which in the rest frame is just its mass.  Hence (from Eq.~(\ref{time})) ${\rm Re}\,\mu = m$, and we conclude that
\begin{equation}
\boxed{\mu = m - \frac{i}{2}\Gamma}\;.
\label{mGamma}
\end{equation}
The same result has been obtained in Ref.~\cite{BS} by different means, also using $\Gamma = 1/\tau$. Eq.~(\ref{mGamma}) is also used to define the mass and width of hadronic resonances (see Sec.~50 of Ref.~\cite{PDG}), and is used, in matrix form, in the analysis of neutral meson oscillations (see Sec.~13 of Ref.~\cite{PDG}). Eq.~(\ref{mGamma}) provides an unambiguous decomposition of $\mu$ into physically meaningful quantities. Note that nowhere did we make a nonrelativistic approximation.

Returning to Eq.~(\ref{pole}), and using Eq.~(\ref{mGamma}), we find
\begin{equation}
{\rm Im}\,\mu^2 = -m\Gamma = - {\rm Im}\, \Pi(\mu^2)
\end{equation}
or
\begin{equation}
\Gamma = \frac{1}{m}{\rm Im}\, \Pi(\mu^2)\;.
\label{width}
\end{equation}
This is an implicit formula for the decay rate, since $\mu = m - (i/2)\Gamma$, and can be solved perturbatively. 

The exponential decay of an unstable particle follows from the complex pole in the particle's propagator.  The propagator also has branch-point singularities, which we discuss in Appendix B. These singularities do not affect the exponential decay of the particle. 

Some authors, going back to the original papers on the subject \cite{L,ELOP}, have expressed the opinion that the definition of the mass (and width) of an unstable particle is inherently ambiguous due to the uncertainty principle. A typical argument is that the energy-time uncertainty principle, $\Delta E \Delta t \geq 1$, restricts the accuracy with which the mass can be defined by $\Delta m\sim \Delta E \geq 1/\Delta t = 1/\tau = \Gamma$. This is faulty logic, as $\Delta E$ corresponds not to the uncertainty in the definition of the mass, but to the spread of energies that an unstable particle is produced with, {\it i.e.}, the width. There is no uncertainty in the pole position in the complex energy plane, nor in the mass, which is the real part of the pole position. The same is true of the width.

\newpage

\noindent{\bf Resonance formulae}

\bigskip

In the energy region near the pole at $p_0 = \mu$, the amplitude of Eq.~(\ref{Epole}) can be approximated by neglecting the antiparticle pole at $p_0 = -\mu$,
\begin{equation}
S \sim \frac{1}{E - \mu}
\end{equation}
where $E=p_0$.
The cross section in the resonance region is thus approximately
\begin{equation}
|S|^2 \sim \frac{1}{(E-m)^2 + \Gamma^2/4}
\label{BW}
\end{equation}
which is the well-known Breit-Wigner formula \cite{BW}. The resonance shape has a full width at half maximum of $\Gamma$, which is why $\Gamma$ is called the width. A similar formula may also be obtained from nonrelativistic quantum mechanics, such as in the original paper \cite{BW}. 

Another way to approximate the cross section in the resonance region is to start from Eq.~(\ref{LOprop}). Making the approximation, valid for $\Gamma^2 \ll m^2$,
\begin{equation}
\mu^2 = \left(m-\frac{i}{2}\Gamma\right)^2 \approx m^2-im\Gamma
\label{mGammaapprox}\end{equation}
gives the propagator
\begin{equation}
S \sim \frac{1}{p^2 - m^2 + im\Gamma}
\end{equation}
and hence the cross section
\begin{equation}
|S|^2 \sim \frac{1}{(p^2 - m^2)^2 + m^2\Gamma^2}\;.
\label{RIBW}
\end{equation}
This is often referred to as a relativistic Breit-Wigner formula. The relativistic invariance follows from the inclusion of both particle and antiparticle poles.

The moral is that both versions of the Breit-Wigner formula, Eqs.~(\ref{BW}) and (\ref{RIBW}), are approximations, and should be treated thusly. In addition, the formula $\mu^2 = m^2 - im\Gamma$ [see Eq.~(\ref{mGammaapprox})], which has gained widespread currency, is also an approximation. {\it The precise definition of the mass and width of an unstable particle is $\mu = m - (i/2)\Gamma$ [Eq.~(\ref{mGamma})].}

Inserting Eq.~(\ref{mGamma}) into Eq.~(\ref{LOprop}) without making any approximations gives the propagator
\begin{equation}
S \sim \frac{1}{p^2 - \left(m - \frac{i}{2}\Gamma\right)^2}
\label{propagator}
\end{equation}
and hence the cross section
\begin{equation}
|S|^2 \sim \frac{1}{(p^2 - m^2 + \Gamma^2/4)^2+m^2\Gamma^2}\;.
\label{resonance}
\end{equation}
It may seem unconventional to include the $\Gamma^2/4$ term in the denominator of the above expression, but its presence follows from the precise definition of mass and width provided by Eq.~(\ref{mGamma}). Eq.~(\ref{resonance}) is the relativistic Breit-Wigner formula, Eq.~(\ref{RIBW}), but without taking the limit $\Gamma^2 \ll m^2$.
\newpage

\noindent{\bf Weak boson masses}

\bigskip

The $Z$ and $W$ boson masses have been extracted from experiment by using the resonance formula
\begin{equation}
|S|^2 \sim \frac{1}{(p^2 - M^2)^2 + (p^2{\it \Gamma}/M)^2}
\label{OSBW}
\end{equation}
which corresponds to none of the Breit-Wigner formulae, Eqs.~(\ref{BW}),(\ref{RIBW}), and (\ref{resonance}). This formula is based on the propagator
\begin{equation}
\frac{1}{p^2 - M^2 + ip^2{\it \Gamma}/M}
\label{OSprop}
\end{equation}
which has a troubled history. It was once thought that the mass of an unstable particle could be defined by
\begin{equation}
M^2-m_B^2+{\rm Re}\,\Pi(M^2)=0\;.
\label{Repole}
\end{equation}
in contrast to Eq.~(\ref{pole}). Combining Eqs.~(\ref{bare}) and (\ref{Repole}) yields the momentum-space propagator
\begin{equation}
\frac{1}{p^2 - M^2 + \Pi(p^2) - {\rm Re}\,\Pi(M^2)}
\approx \frac{1}{p^2 - M^2 + i{\rm Im}\,\Pi(p^2)}\;
\label{Reprop}
\end{equation}
For a weak boson coupled to massless fermions, the leading-order approximation gives ${\rm Im}\,\Pi(p^2) = p^2{\it \Gamma}/M$, where ${\it \Gamma}$ is evaluated at tree level. Inserting this into Eq.~(\ref{Reprop}) gives the propagator of Eq.~(\ref{OSprop}). This is sometimes called the on-shell scheme for unstable particles.

The trouble with this scheme are not the aforementioned approximations, which can be improved; it is that the mass $M$ is gauge dependent \cite{Sirlin} and thus unphysical. Despite this, the parameterization of the resonance region of Eq.~(\ref{OSBW}) has persisted, although it has no good theoretical justification.

Despite appearances, the propagator of Eq.~(\ref{OSprop}) is equivalent to that of Eq.~(\ref{propagator}), and gives exactly the same resonance shape. Both propagators have simple poles, and by equating the poles we can find the exact relation between the parameters $M$ and ${\it \Gamma}$ and the physical mass and width of the particle,
\begin{equation}
m = M \left(\frac{r+1}{2r^2}\right)^{\frac{1}{2}}\approx M\left(1-\frac{3}{8}\left(\frac{{\it \Gamma}}{M}\right)^2\right)
\label{m}
\end{equation}
\begin{equation}
\Gamma = 2M \left(\frac{r-1}{2r^2}\right)^{\frac{1}{2}}\approx {\it \Gamma}\left(1-\frac{5}{8}\left(\frac{{\it \Gamma}}{M}\right)^2\right)
\label{Gamma}
\end{equation}
where
\begin{equation}
r=\left(1+\left(\frac{{\it \Gamma}}{M}\right)^2\right)^{\frac{1}{2}}\;.
\end{equation}

The world-average values $M_Z = 91.1876 \pm 0.0021\; {\rm GeV}$ and ${\it \Gamma}_Z = 2.4952 \pm 0.0023\; {\rm GeV}$ \cite{PDG} can be used to derive the physical $Z$ boson mass and width from Eqs.~(\ref{m}) and (\ref{Gamma}),
\begin{eqnarray}
m_Z = 91.1620 \pm 0.0021\; {\rm GeV}\\
\Gamma_Z = 2.4940 \pm 0.0023\; {\rm GeV}\;.
\end{eqnarray}
The physical $Z$ boson mass is about 26 MeV less than the parameter $M_Z$, a result we derived long ago \cite{WV}; this is about ten times greater than the uncertainty in the mass. The $Z$ boson width is the same as the parameter ${\it \Gamma}_Z$ within the uncertainty. This yields a $Z$ boson lifetime of $\tau_Z = 2.6391 \pm 0.0024 \times 10^{-25}\; {\rm s}$.

The world-average values $M_W = 80.379 \pm 0.012\; {\rm GeV}$ and ${\it \Gamma}_W = 2.085 \pm 0.042\; {\rm GeV}$ \cite{PDG} yield the physical $W$ boson mass and width
\begin{eqnarray}
m_W = 80.359 \pm 0.012\; {\rm GeV} \\
\Gamma_W = 2.084 \pm 0.042\; {\rm GeV}\;.
\end{eqnarray}
The physical $W$ boson mass is about 20 MeV less than the parameter $M_W$, which is nearly twice the uncertainty in the mass. The $W$ boson width is the same as the parameter ${\it \Gamma}_W$ within the uncertainty, and yields $\tau _W = 3.158 \pm 0.064 \times 10^{-25} \; {\rm s}$.

Going forward, it would be simpler to use the propagator of Eq.~(\ref{propagator}) to model the resonance region directly in terms of the physical mass and width, rather than the poorly justified propagator of Eq.~(\ref{OSprop}). 

The top-quark mass and width are also extracted from experiment using the parameterization of Eq.~(\ref{OSBW}). The world-average values are $M_t = 172.76 \pm 0.30\; {\rm GeV}$ and ${\it \Gamma}_t = 1.42 + 0.19 - 0.15\; {\rm GeV}$ \cite{PDG}. The width is sufficiently narrow that these values are equal to the physical top quark mass and width well within the uncertainties. In addition, the physical top quark mass is ambiguous by an amount of order $\Lambda_{QCD}\sim 200\; {\rm MeV}$ \cite{SW}, just like any quark. 

The Higgs boson width is expected to be so narrow ($\sim 4$ MeV) that the difference between the physical mass and the parameter $M_H$ is negligible.

\newpage

{\bf Discussion}
 
\bigskip
 
The relation $\mu = m - (i/2)\Gamma$ [Eq.~(\ref{mGamma})] is used to define the mass and width of hadronic resonances (see Sec.\ 50 of Ref.\ \cite{PDG}) and is also used, in matrix form, in the analysis of neutral meson oscillations (see Sec.~13 of Ref.~\cite{PDG}). In the electroweak sector, an alternative decomposition of the pole position in terms of mass and width, $\mu^2 = m^2 - im\Gamma$ [Eq.~(\ref{mGammaapprox})], has gained widespread currency. How did this come to pass? Up until the discovery of the $W$ and $Z$ bosons, all particles that decayed via the electroweak interactions had widths that were many orders of magnitude less than their mass; the muon, for example, has $\Gamma_\mu/m_\mu = 3 \times 10^{-17}$. The formula $\mu^2 = m^2 - im\Gamma$ is therefore an excellent approximation to $\mu = m - (i/2)\Gamma$. Over time it was forgotten that $\mu^2 = m^2 - im\Gamma$ is an approximation, and it was taken to be exact. This approximation is sufficiently accurate for all particles that decay via the electroweak interaction except for the $W$ and $Z$ bosons, where $\Gamma_Z^2/m_Z^2\sim \Gamma_W^2/m_W^2\sim 0.0007$ is greater than the accuracy of the measurements of the masses ($\Delta m_Z/m_Z \sim 0.00002$, $\Delta m_W/m_W \sim 0.0002$).
 
The expression $\mu^2 = m^2 -im\Gamma$ is often used as an intermediate step for the $W$ and $Z$ boson masses in precision electroweak analyses \cite{F}. One could even view this as an alternative definition of $m$ and $\Gamma$ \cite{S2}. There are several shortcomings to this argument. If we define $m$ and $\Gamma$ this way, then $\Gamma$ is not the decay rate and $m$ is not the oscillation frequency. We have a perfectly good way of defining mass and width precisely, $\mu = m - (i/2)\Gamma$, so why would we want to introduce an alternative definition? Why would we introduce an alternative definition for the $W$ and $Z$ bosons but not for hadronic resonances? And why this particular alternative definition, whose only motivation is that it is approximately equal to the precise definition? Indeed, the current standard definition of the $W$ and $Z$ boson masses is yet another alternative definition, $\mu^2 = M^2/(1+i
\frac{\it \Gamma}{M})$, where $M$ and $\it \Gamma$ are related to $m$ and $\Gamma$ in a complicated way [see Eqs.~(\ref{m}) and (\ref{Gamma})]. These alternative definitions should be regarded as useful parameters, not as the fundamental mass and width of the $W$ and $Z$ bosons. Just as the electron has a unique mass, so do the $W$ and $Z$ bosons. 

Our goal in physics is not only to make precise measurements, but to also be precise about our concepts. With $\mu = m - (i/2)\Gamma$ as the definition of the mass and width, the time dependence of an unstable particle at rest is given by 
\begin{equation}
e^{-imt}e^{-\frac{1}{2}\Gamma t} 
\end{equation}
as desired, while any other definition leads to a bizarre time dependence. For example, the definition $\mu^2 = m^2 -im\Gamma$ yields the time dependence
\begin{equation}
e^{-i\frac{1}{\sqrt 2}m\left(1+ \sqrt{1+\Gamma^2/m^2}\right)^{1/ 2}t}
e^{-\frac{1}{\sqrt 2}\Gamma \left(1 + \sqrt{1+\Gamma^2/m^2}\right)^{-1/ 2}t}\;.
\end{equation}
With this definition it is evident that $\Gamma$ is not the decay rate and $m$ is not the mass.  An even more bizarre time dependence is arrived at using the current standard definition of the $W$ and $Z$ boson masses,
\begin{equation}
e^{-i\frac{1}{\sqrt 2}M\left(1+ \sqrt{1+\it \Gamma^2/M^2}\right)^{1/ 2}\left(1+\it \Gamma^2/M^2\right)^{-1/2}t} e^{-\frac{1}{\sqrt 2}\it M\left(\sqrt{1+\it \Gamma^2/M^2}-1\right)^{1/ 2}\left(1+\it \Gamma^2/M^2\right)^{-1/2}t}\;,
\end{equation}
again demonstrating that $\it \Gamma$ is not the decay rate and $M$ is not the mass.
\bigskip

\noindent{\bf Acknowledgements}

\bigskip

I am grateful for conversations with Christof Hanhart, Andreas Kronfeld, and Jeff Richman.

\newpage

\noindent{\bf Appendix A}

\bigskip

In Eq.~(\ref{FT}), we evaluated the Fourier transform of the scattering amplitude in the rest frame of the unstable particle. If we instead evaluate it in a frame where the particle has three-momentum ${\bf p}$, we obtain [using $\mu = m - (i/2)\Gamma$]
\begin{equation}
S \sim \int_{-\infty}^\infty dp_0\;  \frac{e^{-ip_0t}}{\left(p_0 - \sqrt{{\bf p}^2+\mu^2}\right)\left(p_0 + \sqrt{{\bf p}^2+\mu^2}\right)} \sim e^{-i\sqrt{{\bf p}^2+\mu^2}t} \sim e^{-iEt}e^{-(m/E)(\Gamma/2)t}
\end{equation}
where the energy $E$, identified by the oscillatory time dependence of the amplitude, is 
\begin{equation}
E= \frac{1}{\sqrt 2}\left({\bf p}^2+m^2-\Gamma^2/4+\sqrt{\left({\bf p}^2+m^2-\Gamma^2/4\right)^2+m^2\Gamma^2}\right)^{1/2}\;,
\end{equation}
which reduces to $m$ when ${\bf p}=0$. As expected, the decay rate is reduced by a factor of $m/E$, which is the inverse of the usual time-dilation factor. This calculation generalizes that of Ref.~\cite{B}, which is done using the approximation $\Gamma \ll m$.

\bigskip

\noindent{\bf Appendix B}

\bigskip

The momentum-space propagator of an unstable particle has the analytic structure in the complex $p^2$ plane shown in Fig.~3 (see Sec.~50 of Ref.~\cite{PDG}).  There is a branch point at the lowest threshold, with a branch cut extending to infinity along the positive real $p^2$ axis.  There is a new branch point and branch cut for each new threshold, but we show only the lowest threshold for clarity of presentation. The physical axis lies on the first sheet, just above the branch cut. The pole in the propagator lies on the second sheet at $p^2 = \mu^2$, which is accessed by passing downward through the branch cut from the physical axis.  Since the self energy $\Pi(p^2)$ satisfies the Schwarz reflection principle, there is also a pole on the second sheet at $p^2 = \mu^{*2}$.  This pole is much further from the physical axis than the pole at $p^2=\mu^2$. 

The analytic structure of the momentum-space propagator in the complex energy ($p_0$) plane, with ${\bf p} = 0$, is shown in Fig.~4. There is a branch point at the lowest threshold on the positive real axis as well as on the negative real axis.  There is a particle pole on the second sheet below the branch cut on the positive real axis, and an antiparticle pole on the second sheet above the branch cut on the negative real axis.  As in the $p^2$ plane, there are also poles at the complex conjugate positions, far from the physical axis.  Also shown is the contour of integration for the Fourier transform of the propagator.

The Fourier transform can be performed by closing the contour on the first sheet, as shown in Fig.~5. Since the poles lies on the second sheet, they are not shown in the figure, and are not enclosed by the contour.  Let the momentum-space propagator be denoted by 
\begin{equation}
G(p_0) = \frac{iZ^{-1}}{p_0^2 - \mu^2 + \Pi(p_0^2) - \Pi(\mu^2)+i\epsilon} \;.
\end{equation}
Then, for $t>0$, the contour integration gives
\begin{equation}
\int_{-\infty}^\infty dp_0\; e^{-ip_0t}G(p_0) = \int_{thr}^\infty dp_0\; e^{-ip_0t}\;{\rm Disc}\;G(p_0)  
\label{Disc}
\end{equation}
where ${\rm Disc}\;G(p_0)$ is the discontinuity of $G(p_0)$ across the branch cut running from threshold to infinity. Let's compare this with the Fourier transform of the K\"allen-Lehmann representation of the propagator (see Sec.~10.7 of Ref.~\cite{W}), 
\begin{equation}
\int_{-\infty}^\infty dp_0\; e^{-ip_0t}G(p_0) = \int_{-\infty}^\infty dp_0\; e^{-ip_0t} \int_{thr^2}^\infty \frac{dM^2}{2\pi}\rho(M^2)\frac{i}{p_0^2-M^2+i\epsilon} 
\end{equation}
where $\rho(M^2)$ is the spectral density. For $t>0$ we close the $p_0$ contour in the lower half plane and pick up the residue of the pole at $p_0 = M-i\epsilon$, yielding
\begin{equation}
\int_{-\infty}^\infty dp_0\; e^{-ip_0t}G(p_0) = \int_{thr}^\infty dM\; e^{-iMt}\rho(M^2) 
\end{equation}
Comparing with Eq.~(\ref{Disc}) we conclude
\begin{equation}
\rho(p_0^2) = {\rm Disc}\;G(p_0) 
\end{equation}
Note that the spectral function does not have a delta function contribution from the unstable particle, as it would from a stable particle, because the unstable particle is not part of the spectrum of asymptotic states.

To isolate the contribution from the unstable particle, it is necessary to close the contour on the second sheet, as shown in Fig.~6, rather than on the first sheet as was done above. One finds
\begin{equation}
\int_{-\infty}^\infty dp_0\; e^{-ip_0t}G(p_0) 
=-2\pi i (e^{-i\mu t}+e^{i\mu^* t} ) - \int_{thr}^{-\infty} dp_0\; e^{-ip_0t}\;{\rm Disc}\;G(p_0)
\label{secondsheet}\end{equation}
where the discontinuity is between $G(p_0)$ on the first and second sheets. This alternative evaluation of the Fourier transform makes the exponential decay associated with the unstable particle pole explicit. The integral on the right-hand side also contributes to the time dependence of the Fourier transform of the propagator, but it does not yield exponential decay.

There may be other branch cuts that contribute to the time dependence of the Fourier transform of the propagator, but these also do not yield exponential decay. For example, consider the case of an unstable particle with electric or color charge. The self-energy has a contribution from the unstable particle emitting and then reabsorbing a virtual photon or gluon. The self-energy at one loop is proportional to \cite{DDRW2}
\begin{equation}
\Pi(p^2) \sim (\mu^2/p^2-1)\ln(1-p^2/\mu^2)+\cdots
\end{equation}
which has a logarithmic branch point at $p^2=\mu^2$, corresponding to an infrared singularity due to soft virtual photons or gluons. The branch-point singularity is removed via field renormalization, which sets the residue of the pole to unity. The field renormalization constant $Z$ is thus infrared divergent, but this divergence cancels against other infrared divergences in physical quantities in the usual way (see Secs.~13.1 - 13.3 of Ref.~\cite{W}). The resulting analytic structure in the $p_0$ plane is shown in Fig.~7. The contour on the second sheet must avoid the branch cuts that emanate from the poles, but the small circles around the poles still pick up the residues of the poles, and one again arrives at Eq.~(\ref{secondsheet}), but with additional integrals on the right-hand side from the discontinuities across the photon or gluon branch cuts.

\begin{figure}
\centering
\includegraphics[width=\textwidth]{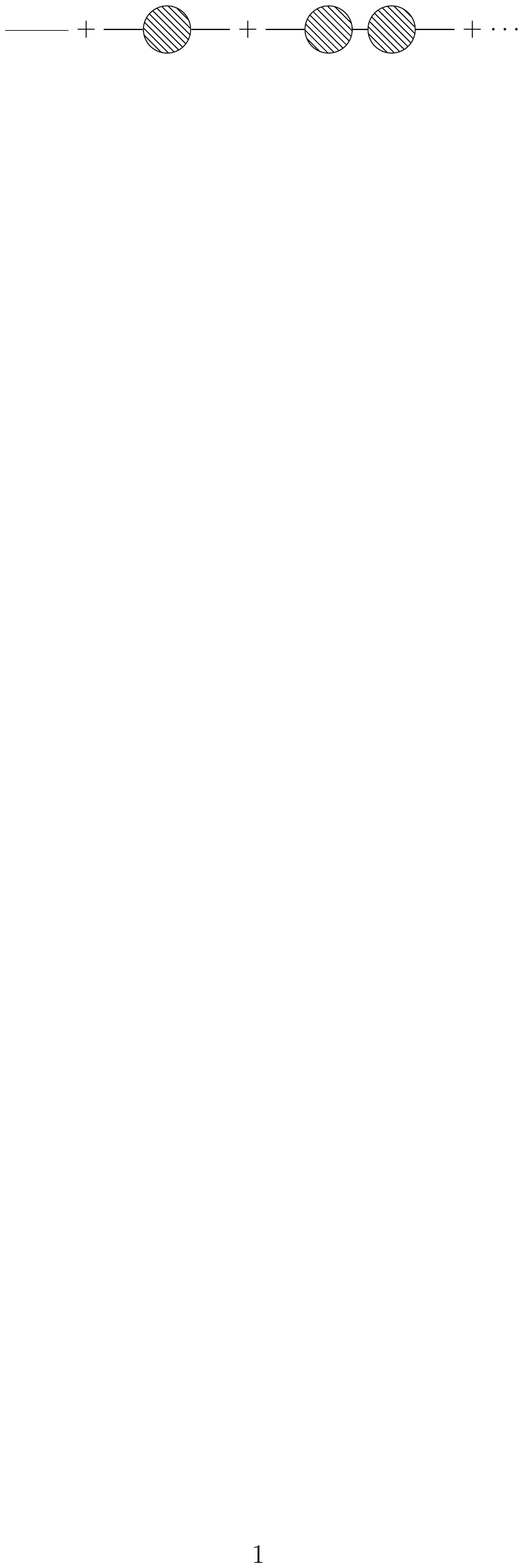}
\caption{Self-energy corrections to the propagator.}
\end{figure}

\begin{figure}
\centering
\includegraphics[width=\textwidth]{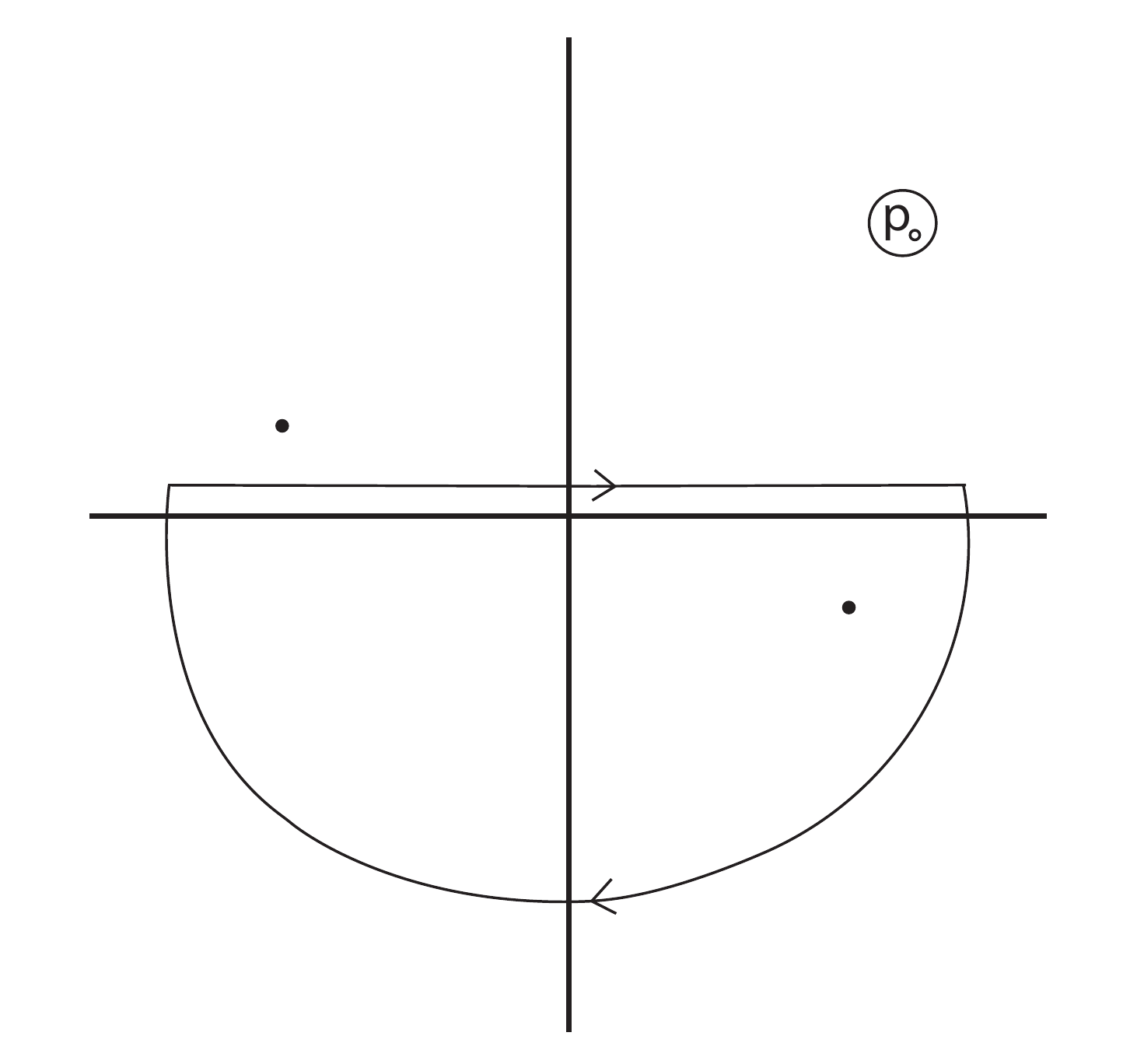}
\caption{Evaluating the Fourier transform of the propagator by closing the contour in the lower-half energy plane (for $t>0)$.}
\end{figure}

\begin{figure}
\centering
\includegraphics[width=\textwidth]{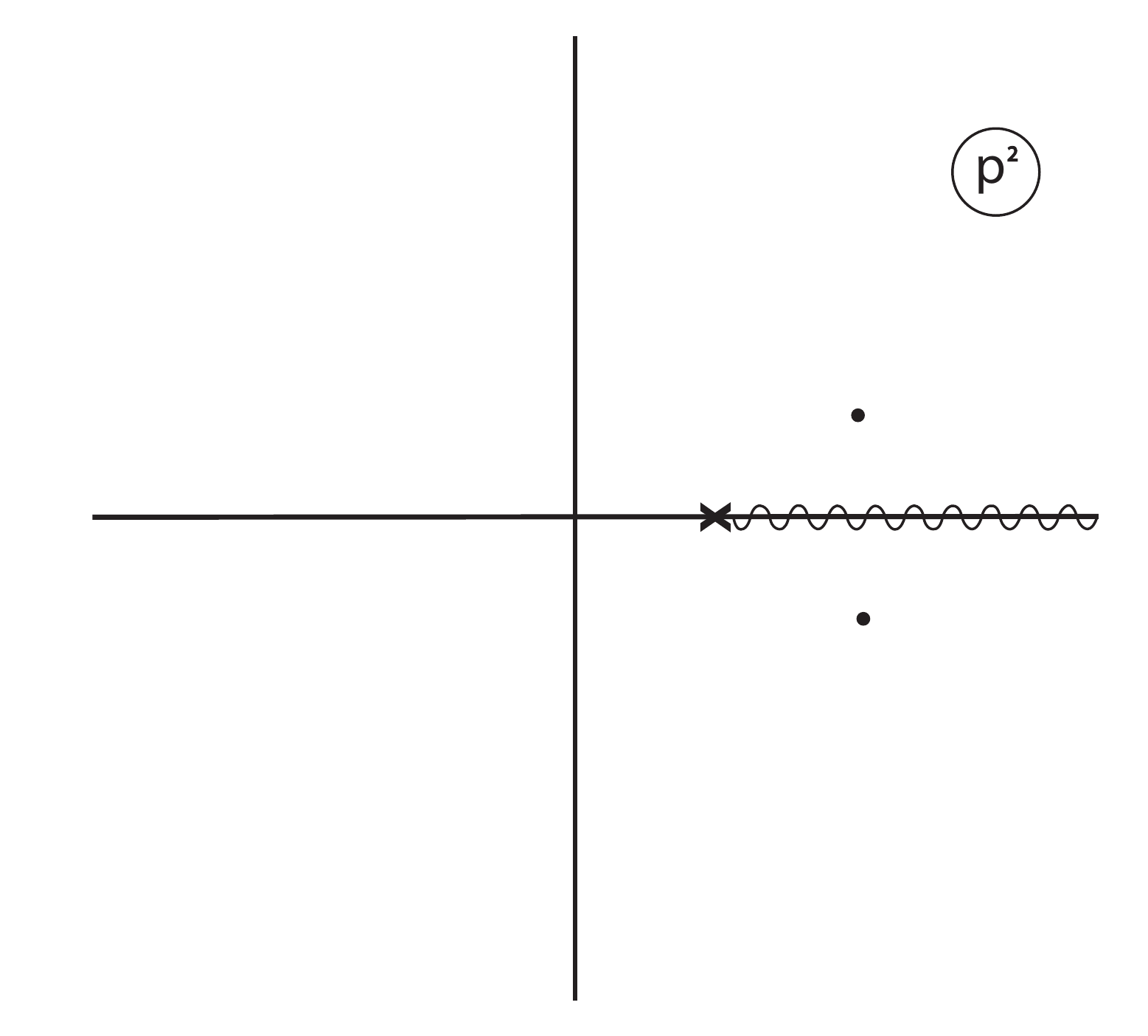}
\caption{Analytic structure of the propagator in the complex $p^2$ plane. The poles lie on the second sheet.}
\end{figure}

\begin{figure}
\centering
\includegraphics[width=\textwidth]{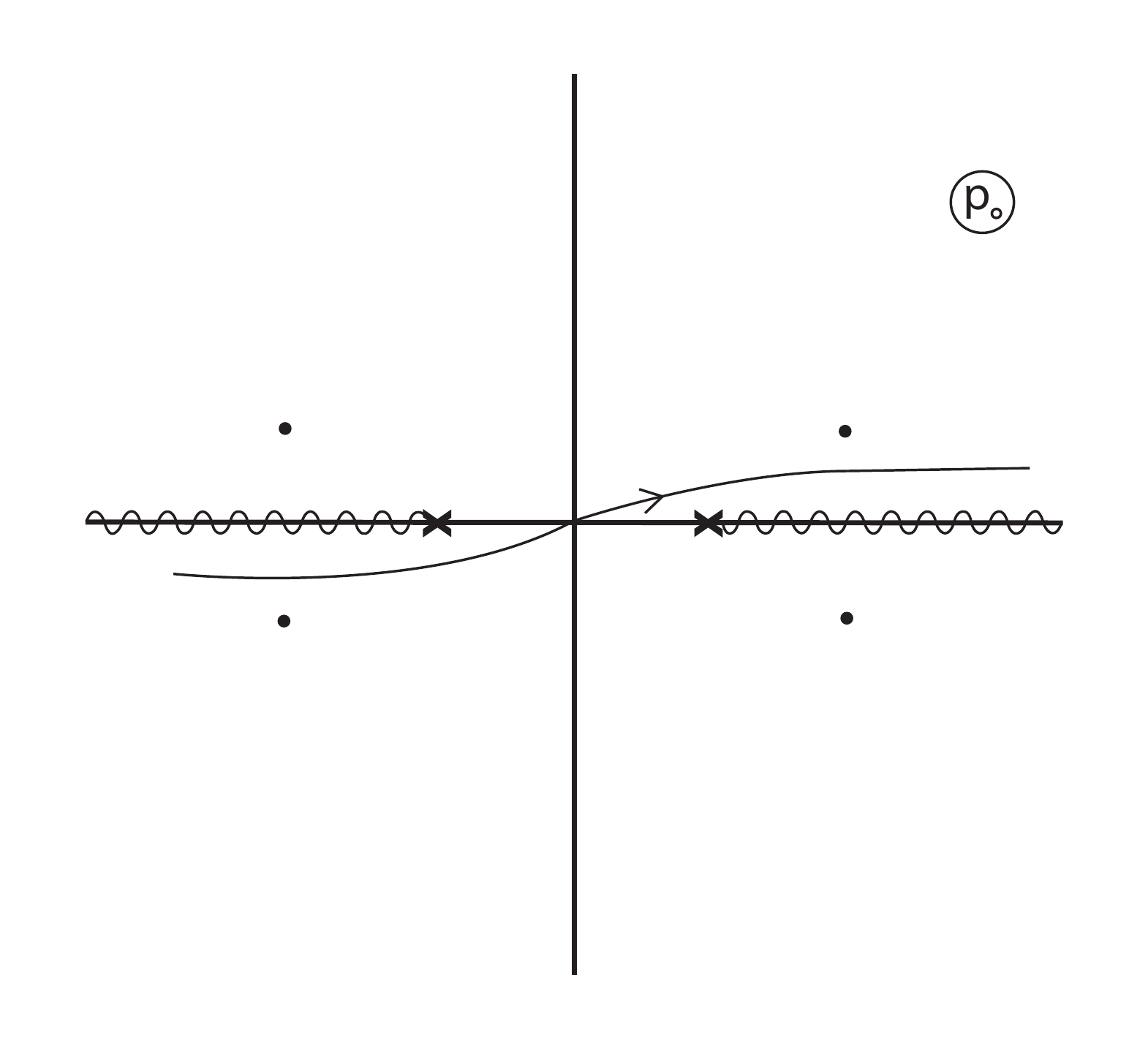}
\caption{Analytic structure of the propagator in the complex energy plane. The poles lie on the second sheet. The integration contour for the Fourier transform is also shown.}
\end{figure}

\begin{figure}
\centering
\includegraphics[width=\textwidth]{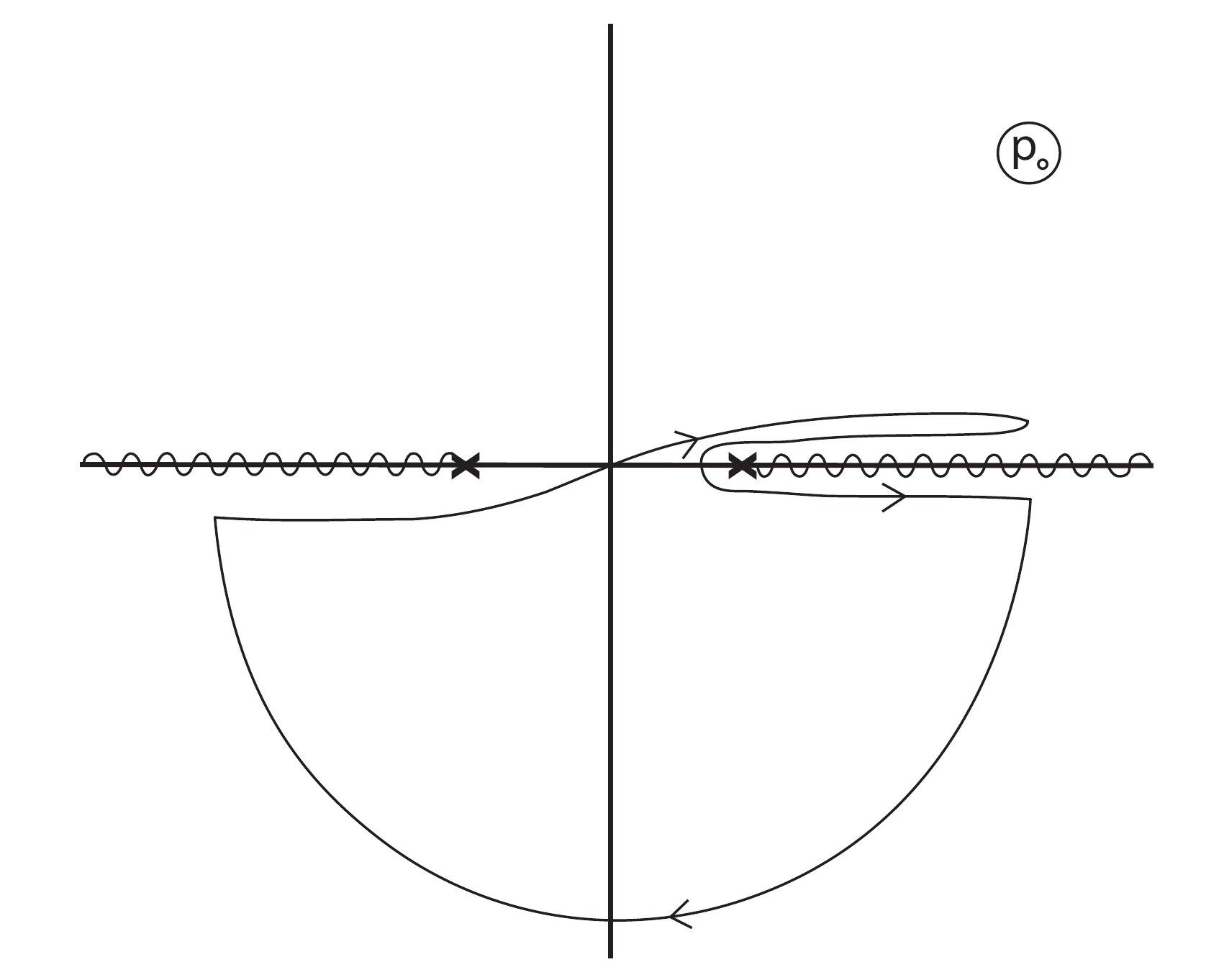}
\caption{Evaluating the Fourier transform of the propagator by closing the contour in the lower-half energy plane (for $t>0)$ on the first sheet.}
\end{figure}

\begin{figure}
\centering
\includegraphics[width=\textwidth]{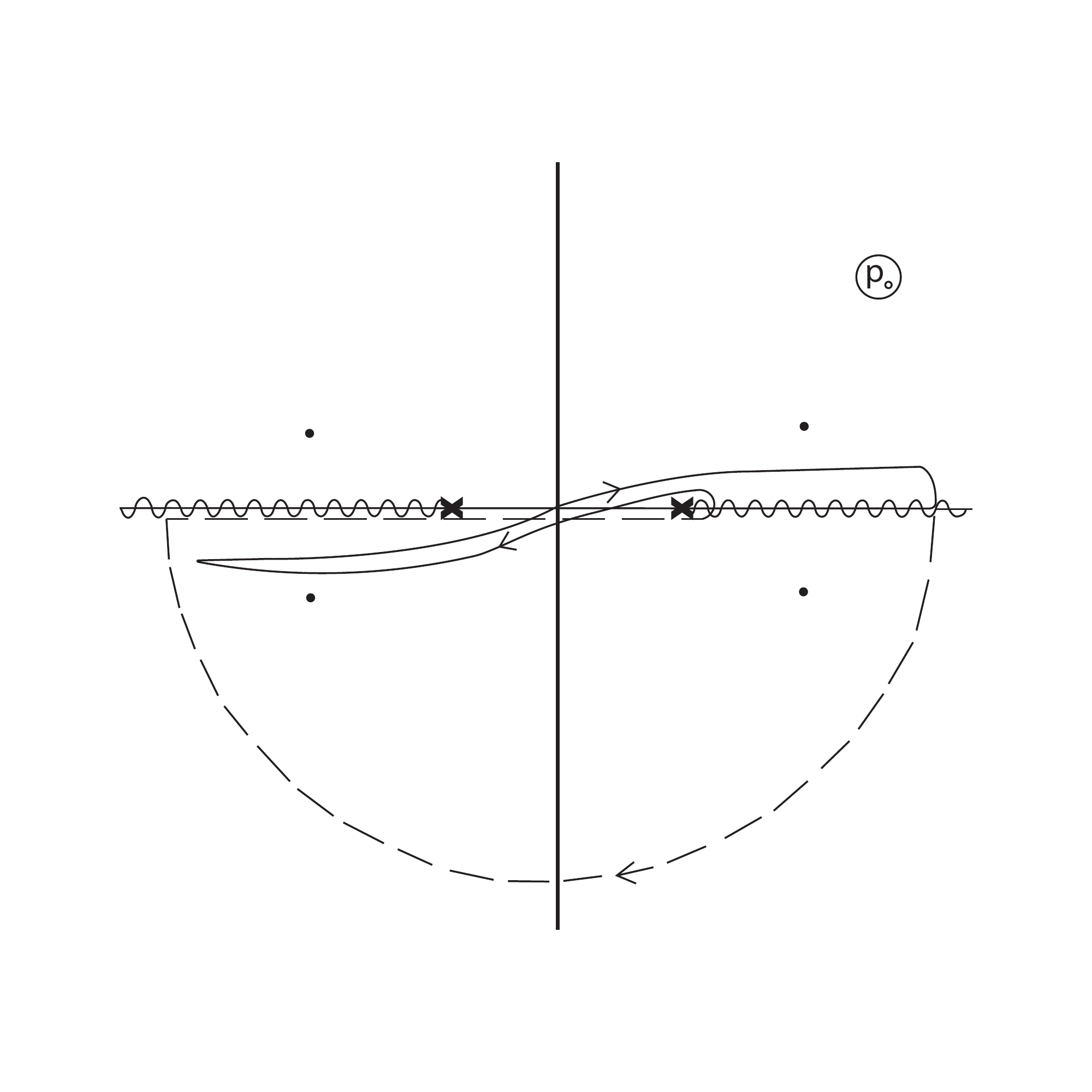}
\caption{Evaluating the Fourier transform of the propagator by closing the contour in the lower-half energy plane (for $t>0)$ on the second sheet.}
\end{figure}

\begin{figure}
\centering
\includegraphics[width=\textwidth]{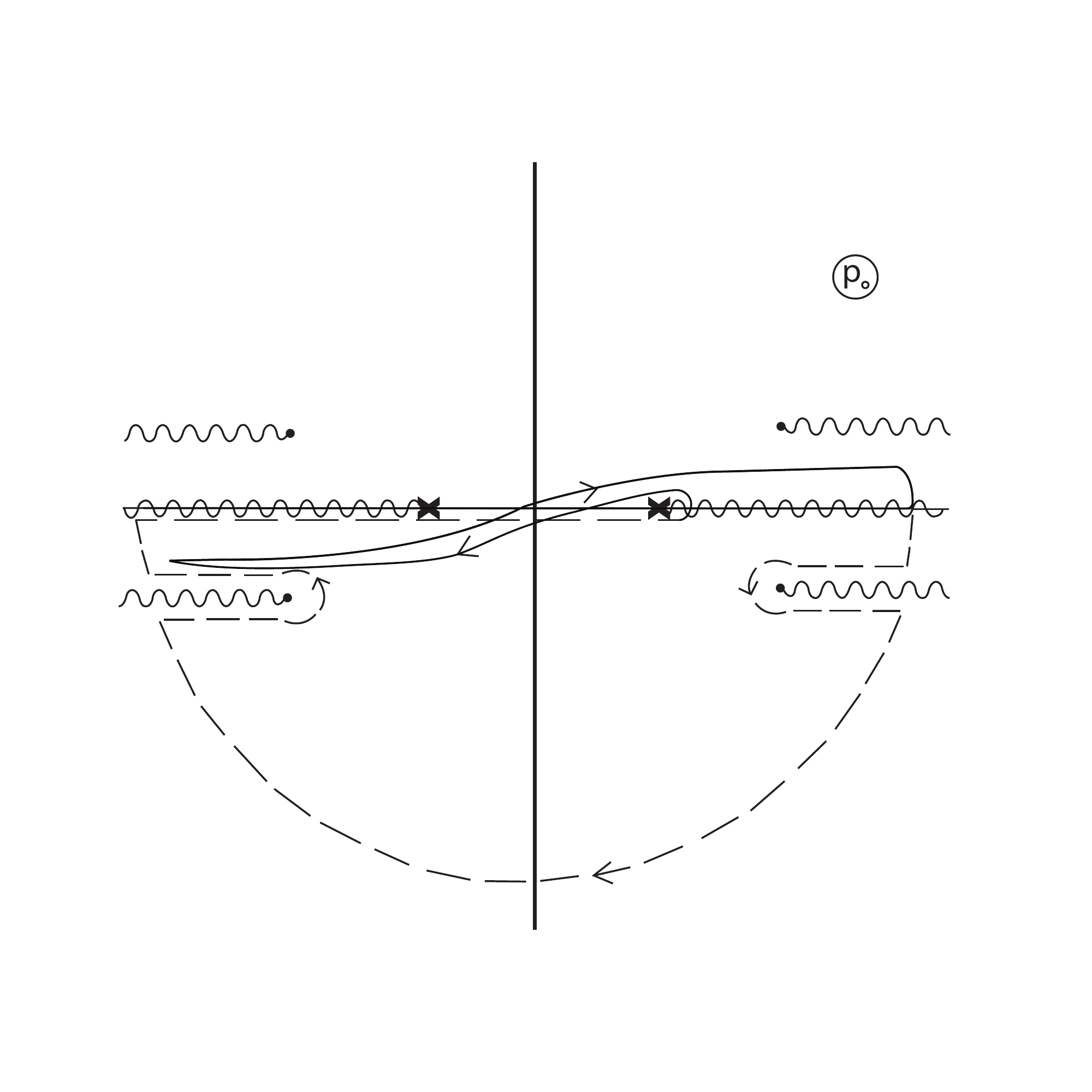}
\caption{Same as Fig.~6, but with additional branch cuts from a photon or gluon being emitted and reabsorbed by the unstable particle.}
\end{figure}


\newpage

\begin{thebibliography}{100}

\bibitem{W} S.~Weinberg, {\sl The Quantum Theory of Fields, Vol.~I} (Cambridge University Press, 1995).

\bibitem{DDRW} A.~Denner, S.~Dittmaier, M.~Roth, and L.~Wieders, Nucl.\ Phys.\  {\bf B724}, 247 (2005).
    
\bibitem{S} R.~Stuart, Phys.\ Lett.\ B {\bf 262}, 113 (1991).

\bibitem{K} A.~S.~Kronfeld, Phys.\ Rev.\ D {\bf 58}, 051501 (1998).

\bibitem{GG} P.~Gambino and P.~A.~Grassi, Phys.\ Rev.\ D {\bf 62}, 076002 (2000).

\bibitem{B} L.~Brown, {\sl Quantum Field Theory} (Cambridge University Press, 1992), Sec.~6.3.

\bibitem{BS} A.~Bohm and Y.~Sato, Phys.\ Rev.\ D {\bf 71}, 085018 (2005).

\bibitem{PDG} R.~L.~Workman {\it et al.} (Particle Data Group), Prog.\ Theor.\ Exp.\ Phys.\ {\bf 2022}, 083C01 (2022).

\bibitem{L} M.~L\'evy, Nuovo Cimento {\bf 13}, 115 (1959).

\bibitem{ELOP} R.~J.~Eden, P.~V.~Landshoff, D.~I.~Olive, and J.~C.~Polkinghorne, {\sl The Analytic S-Matrix} (Cambridge University Press, 1966), Sec.~4.9.

\bibitem{BW} G.~Breit and E.~P.~Wigner, Phys.\ Rev.\ {\bf 49}, 519 (1936).

\bibitem{Sirlin} A.~Sirlin, Phys.\ Rev.\ Lett.\ {\bf 67}, 2127 (1991).

\bibitem{WV} S.~Willenbrock and G.~Valencia, Phys.\ Lett.\ B {\bf 259}, 373 (1991).

\bibitem{SW} M.~Smith and S.~Willenbrock, Phys.\ Rev.\ Lett.\ {\bf 79}, 3825 (1997).

\bibitem{F} A.~Freitas, Precision Tests of the Standard Model, in {\sl Proceedings of the Theoretical Advanced Studies Institute 2020} (PoS TASI 2020), Vol.~388, 005 (2021).

\bibitem{S2} R.~Stuart, Phys.\ Rev.\ Lett.\ {\bf 70}, 3193 (1993).

\bibitem{DDRW2} A.~Denner, S.~Dittmaier, M.~Roth, and L.~Wieders, Nucl.\ Phys.\  {\bf B854}, 504 (2012).

\end{thebibliography}
\end{document}